\documentclass[aps,prl,reprint,showpacs,superscriptaddress]{revtex4-1}
\usepackage{amsfonts,mathrsfs,amsmath,amsthm,amssymb,bbold}
\usepackage{epsfig}
\usepackage{bm,mathrsfs}
\usepackage{lineno,hyperref}
\usepackage{amsmath}
\usepackage{color}
\usepackage{xcolor,colortbl} 
\usepackage{soul}
\usepackage{afterpage}
\usepackage{capt-of}
\usepackage{makecell}
\usepackage{graphicx}
\setcellgapes{4pt}

\newcolumntype{P}[1]{>{\centering\arraybackslash}p{#1}}

\usepackage{amsmath} 

\begin{document}
\sloppy
\clearpage

\title[mode = title]{Skyrmions in van der Waals centrosymmetric materials with Dzyaloshinskii--Moriya interactions}

%
\author{Hung Ba Tran}
\address{Laboratory for Materials and Structures, Institute of Innovative Research, Tokyo Institute of Technology, Midori-ku, Yokohama 226-8503, Japan}
\address{Quemix Inc., Taiyo Life Nihombashi Building, 2-11-2, Nihombashi, Chuo-ku, Tokyo 103-0027, Japan}

\author{Yu-ichiro Matsushita}
\address{Laboratory for Materials and Structures, Institute of Innovative Research, Tokyo Institute of Technology, Midori-ku, Yokohama 226-8503, Japan}
\address{Quemix Inc., Taiyo Life Nihombashi Building, 2-11-2, Nihombashi, Chuo-ku, Tokyo 103-0027, Japan}
\address{Quantum Material and Applications Research Center, National Institutes for Quantum Science and Technology, 2-12-1, Ookayama, Meguro-ku, Tokyo 152-8552, Japan}

\date{\today}

\begin{abstract}
Skyrmions can appear in non-centrosymmetric material because of the non--vanishing Dzyaloshinskii--Moriya interactions (DMI). In this paper, we study the magnetic properties of the rhombohedral MX$_{3}$ (M: V, Cr, Mn, and Fe. X: Cl, Br, and I) van der Waals materials with centrosymmetric lattice by combining first--principles calculations and Monte Carlo simulations. We found that the Dzyaloshinskii--Moriya vector acting between the second nearest neighbor sites of the intra--layer is non--zero and quite large even in MX$_{3}$ due to the breaking of the local inversion symmetry. We have found that the large DMI causes nano--scale magnetic vortexes, so--called skyrmions in MX$_{3}$. The diameter of skyrmions, e.g., in CrCl$_{3}$, is small, i.e., 2 nm, which is promising for the application of high--density magnetic memory devices. We have also found that not only conventional skyrmions in CrCl$_{3}$ and VCl$_{3}$ but also antiferromagnetic skyrmions in FeCl$_{3}$ and meron in MnCl$_{3}$ appear. Furthermore, we have found that the skyrmions in ClCl$_{3}$ and VCl$_{3}$ have a different helicity, which shows the possibility of controlling the helicity by electron/hole doping in MX$_{3}$ materials. Van der Waals materials, which have a great advantage of high degrees of freedom in structures such as heterostructures and twisted structures, exhibit high potential as skyrmion materials.  
\end{abstract}

\maketitle
Skyrmion, a tiny magnetic swirl, is promising for future spintronic applications such as race track memory devices\cite{Fert2017NRM,Tokura2021CR}. A conventional skyrmion is known to exist in non--centrosymmetric materials due to the antisymmetric exchange Dzyaloshinskii--Moriya interactions (DMI)\cite{Fert2017NRM,Tokura2021CR}. In centrosymmetric materials, the DMI is well known to be zero and can not support the appearance of skyrmion\cite{Moriya1960PR}. Recent experimental studies, however, show that the skyrmion can be realized even in centrosymmetric materials such as triangular lattice\cite{Kurumaji2019S,Khanh2020NN,Takagi2022NC}. It comes from the frustration or oscillation of magnetic exchange coupling constant $J_{\rm ij}$, which is also known as Ruderman--Kittel--Kasuya--Yosida (RKKY) interactions\cite{Kurumaji2019S,Khanh2020NN,Takagi2022NC,Yambe2021SR}. The survivor of a skyrmion in centrosymmetry material brings many advantages such as controllability of helicity and polarity as another degree of freedom, which can be applicable for future computing hardware such as skyrmion qubit in quantum computing or neuromorphic computing\cite{Yambe2021SR,Yao2020NJP,Psaroudaki2021PRL,Jing2022arxiv,Song2020NE}. Explore some other possible mechanism to realize the skyrmion in new class of centrosymmetric materials becomes important tasks for both theoretical and experimental frameworks.   

\begin{figure}[htp] 
\centering
\includegraphics[width=8.6cm]{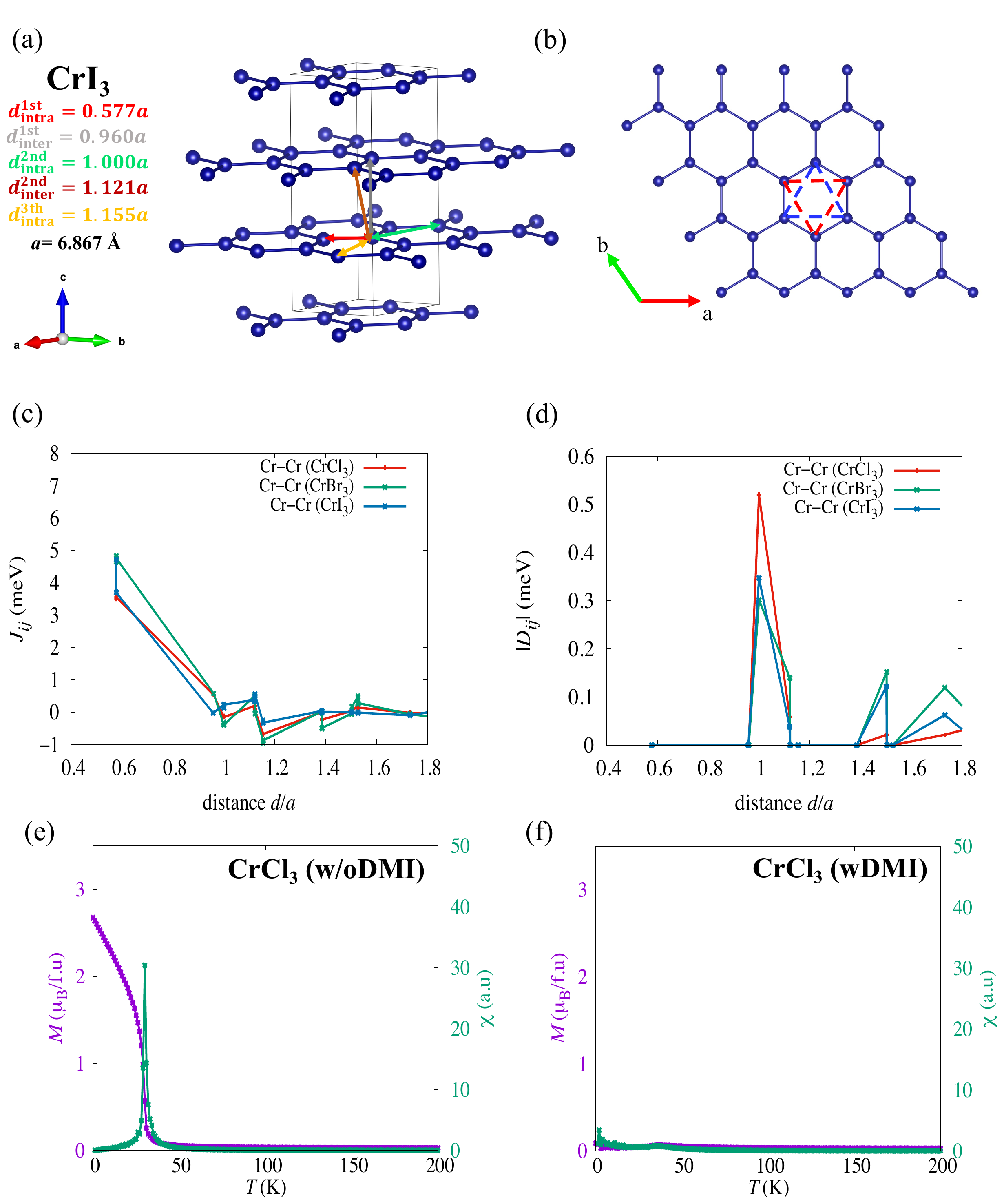} 
\caption{(a) Crystal structure of rhombohedral CrI$_{3}$, where only Cr atom is shown as blue sphere. (b) Layer of honeycomb lattice where two sub--lattice of the second nearest neighbor of intra--layer are indicated as the red and blue dash lines. (c) The magnetic exchange coupling constants of Cr--Cr pairs of CrX$_{3}$ (X: Cl, Br, I) as a function of distance over lattice constant ($d/a$). (d) The length of Dzyaloshinskii--Moriya vector of CrX$_{3}$ as a function of distance over lattice constant ($d/a$). The temperature--dependence of magnetization (purple color) and magnetic susceptibility (green color) of CrCl$_{3}$ without Dzyaloshinskii--Moriya interaction (e) and with Dzyaloshinskii--Moriya interaction (f).} 
\label{FIG1}
\end{figure}

2D materials have a high degree of structural freedom such as heterostructure and twisted structures, and recently offer an interesting arena for future electronic device and various new physics, e.g., superconducting and Fermi velocity modulation in twisted bilayer graphene\cite{Ahn2020NPJ2D,Gibertini2019NN,Cao2018N,Ando2006JPSJ,Koshino2018PRX,Nishi2017PRB,Uchida2014PRB,Dresselhaus2010NL}. The van der Waals magnetic materials such as the ferromagnetic 2D materials CrI$_{3}$ and their family have attracted considerable attention due to their peculiar properties due to their layer structure, which is possible to manufacture a few layers or even monolayer with some new physics properties rather than bulk system such as spin flip and peculiar magnetoresistance properties\cite{Ahn2020NPJ2D,Gibertini2019NN,McGuire2017C}. Bulk CrI$_{3}$ has two crystal structures: the low-temperature phase rhombohedral with space group $R\bar{\rm 3}$, and the high-temperature phase monoclinic with space group $C2/m$\cite{Handy1952JACS}. The low--temperature crystal structure of CrI$_{3}$ ($R\bar{\rm 3}$) is shown as FIG. \ref{FIG1} (a). The crystal structure is centrosymmetric with honeycomb lattice, which leads to the ignorance of DMI. The $ab$ plane of CrI$_{3}$ with two sub--lattice is shown in FIG. \ref{FIG1} (b). The second nearest neighbor of intra--layer has the broken of local inversion symmetry, where the center of this pair does not satisfy the inversion symmetry. It brings a possibility of finite DMI for this pair of atoms in honeycomb lattice even in a centrosymmetric material. It can be the new mechanism to realize the skyrmion lattice in a centrosymmetric material compared with the frustration and RKKY interaction proposed in the previous works\cite{Kurumaji2019S,Khanh2020NN,Takagi2022NC}. The effect of DMI on magnetic properties of MX$_{3}$ (M: V, Cr, Mn, Fe. X: Cl, Br, I) can not be neglected, which can lead to some rich magnetic properties. It motivates us to study the magnetic properties in these compounds with the effect of DMI is taken into account by combining first--principles calculations and Monte Carlo simulations.
 
In this work, we derive Hamiltonian from first-principles calculations, where the essential parts are isotropic exchange, antisymmetric exchange (also known as Dzyaloshinskii--Moriya interaction), magnetic anisotropy, and Zeeman term are taken into account. The Hamiltonian of Heisenberg model is used as\cite{HBT2022PRB,HBT2022AM} 

\begin{equation}
\begin{split}
&H_{\rm Heis}=-\sum_{<ij>}J_{ij}^{m}\overrightarrow{S_{i}}\cdot\overrightarrow{S_{j}} -\sum_{i}k_{\rm u}(\overrightarrow{e_{u}}\cdot\overrightarrow{S_{i}})^{2} \\
& -\sum_{<ij>}\overrightarrow{D_{ij}}\cdot(\overrightarrow{S_{i}} \times \overrightarrow{S_{j}}) -g\mu _{\rm B}\sum_{i}\overrightarrow{H_{\rm ext}}\cdot\overrightarrow{S_{i}},
\end{split}
\label{Eq1}
\end{equation}  

\noindent where $g$ is the $g$-factor, $\mu_{\rm B}$ is the Bohr magneton, $k_{\rm u}$ is the uniaxial anisotropy constant. The first term expresses the exchange interactions between spins at sites \textit{i} and \textit{j}. The spin tends to be parallel to the neighboring site when the magnetic exchange coupling constant $J_{ij}^{m}$ is positive while the spin is antiparallel to the neighboring spin for negative $J_{ij}^{m}$. The magnetic exchange coupling constants can be calculated by using the Liechtenstein formula in density functional theory (DFT)\cite{Liechtenstein1987JMMM,EbertSPRKKR,Ebert2011RPP}. The value of magnetic exchange coupling constants of Cr--Cr pair of CrX$_{3}$ as the function of distance over lattice constant $d/a$ are shown in FIG. 1(c). The first nearest neighbor of intra--layer is a large positive in all compounds, which can lead to the ferromagnetic order of intra--layer. The coupling constants of the first and second nearest neighbor of interlayer in CrCl$_{3}$ are positive. It means the AFM order of interlayer in experimental work can not be realized from the results of DFT\cite{McGuire2017PRM}. The second term in Eq.(\ref{Eq1}) is the interaction between the spin at site \textit{i} and the uniaxial anisotropy, $\overrightarrow{e_{u}}$ being the direction of the easy axis in the case of a positive $k_{\rm u}$. $\overrightarrow{e_{u}}$ is set as out--of--plane ($c$ direction) and $k_{\rm u}$ is evaluated from the value of magnetocrystalline anisotropy energy (MAE) in first-principles calculations\cite{Kresse1996PRB}. The value of MAE in CrCl$_{3}$ is positive (0.034 meV/Cr--atom) and the easy axis is out--of--plane. It is again in contradiction with the in--plane easy plane in reported experimental works, which require the negative value of MAE\cite{McGuire2017PRM}. The third term is the antisymmetric exchange term with $\overrightarrow{D_{ij}}$ being the Dzyaloshinskii--Moriya vector, which can be calculated as the slope of magnon at $\Gamma$ point in the case of spin spiral calculations\cite{EbertSPRKKR,Ebert2011RPP,Mankovsky2017PRB}. The length of the Dzyaloshinskii--Moriya vector of Cr--Cr pair of CrX$_{3}$ as the function of distance over lattice constant $d/a$ is shown as FIG. \ref{FIG1} (d). The length of the Dzyaloshinskii--Moriya vector of the second nearest neighbor of intra--layer is quite large. In the case of CrI$_{3}$, the value of the second nearest neighbor is 0.33 meV, which is in good agreement with the value extracted from the magnon spectrum in experimental work (0.31 meV)\cite{Chen2018PRX}. In addition, the value in CrCl$_{3}$ is comparable with CrI$_{3}$, which can have a significant effect on magnetic properties since the MAE in this material is not so large to suppress the effect of DMI. The final term is Zeeman, which is the effect of the external magnetic field on the spin.

The temperature-dependence of magnetization and magnetic susceptibility of CrCl$_{3}$ are shown in FIG. \ref{FIG1} (e) and (f). In the case that DMI is removed, CrCl$_{3}$ shows a clear normal magnetic transition from ferromagnetic to paramagnetic (FM--PM) in magnetization as well as magnetic susceptibility. It means that the positive interlayer magnetic exchange coupling constants and positive MAE with out--of--plane easy axis in CrCl$_{3}$ can not support the in--plane AFM in the interlayer as reported in experimental work, which required the negative exchange coupling constant of interlayer and in--plane MAE\cite{McGuire2017PRM}. Then, the effect of DMI on magnetization and magnetic susceptibility is taken into account. In the case of CrCl$_{3}$, the MAE (0.034 meV/Cr--atom) is significantly lower than CrBr$_{3}$ (0.187 meV/Cr--atom) and CrI$_{3}$ (0.590 meV/Cr--atom) while the DMI is comparable to these compounds.  It leads to the disappearance of the FM as the ground state and the FM--PM transition in magnetization and magnetic susceptibility. The magnetization at the ground state is approximately equal to zero, and the magnetic susceptibility becomes negligible. It might have the similar features as the reported AFM phase in experimental works. It means that the DMI is the critical factor in determining the ground state of CrCl$_{3}$. Although the FM phase is suppressed with the consideration of DMI, the magnetization appears with the magnetic field applied. Note that the FM--PM transition can survive in CrI$_{3}$ and CrBr$_{3}$ cases even with the DMI since the MAE of them is strong enough to stabilize the FM at low temperature. See Figure 2 in Supplementary materials for the magnetization and magnetic susceptibility dependence on temperature in the other compounds where the Dzyaloshinskii--Moriya interactions are included.

\begin{figure}[htp] 
\centering
\includegraphics[width=8.6cm]{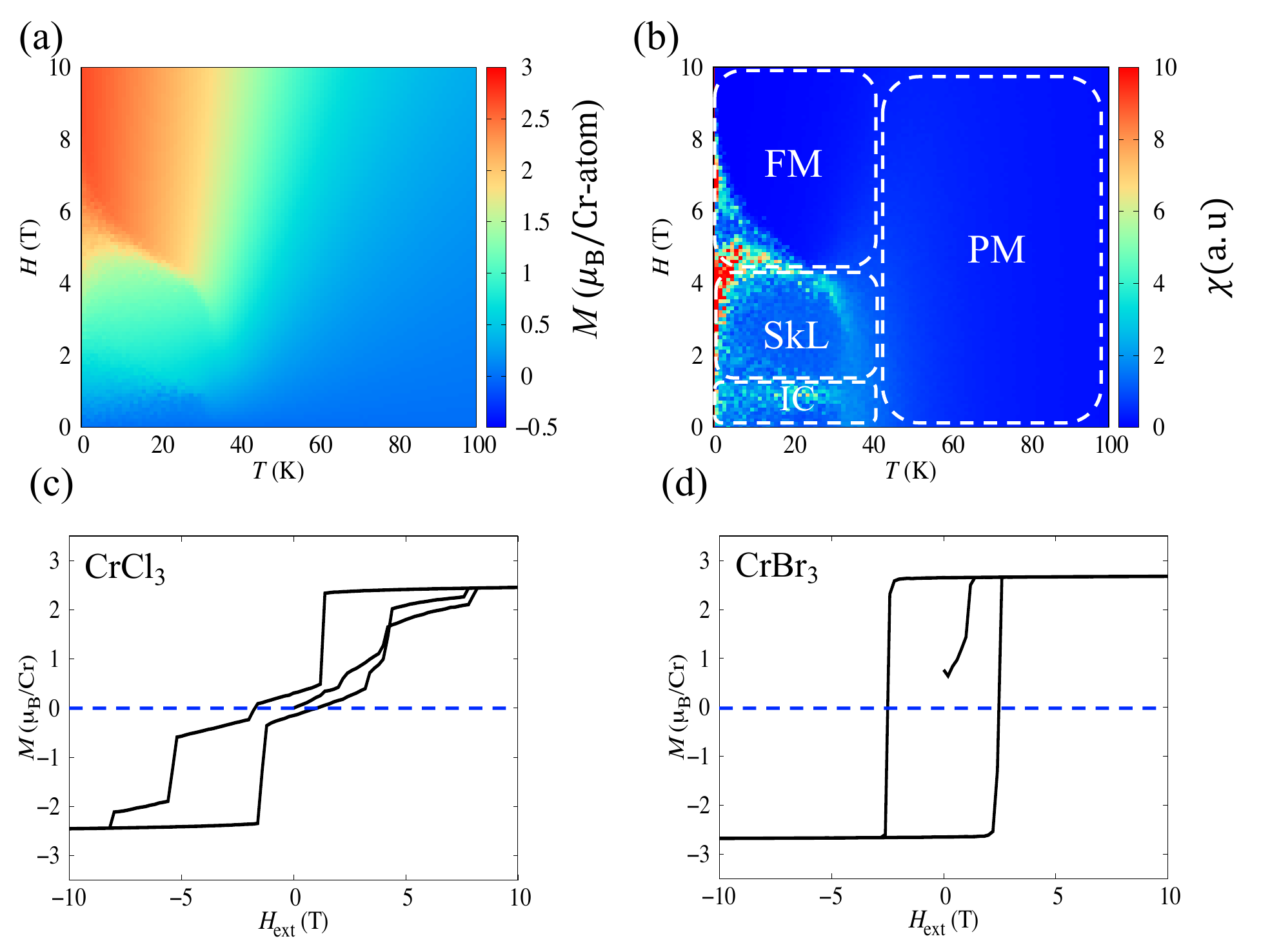} 
\caption{(a) The magnetization of CrCl$_{3}$ as a function of temperature (horizontal axis) and the external magnetic field (vertical axis). (b) The magnetic susceptibility of CrCl$_{3}$ as the function of temperature and the external magnetic field, where the magnetic phase is also specified. The hysteresis loop of CrCl$_{3}$ (c) and CrBr$_{3}$ (d) at 10 K and low frequency of field scan.} 
\label{FIG2}
\end{figure}

The magnetization of CrCl$_{3}$ as the function of temperature and external magnetic field along $z$ direction is shown in FIG. \ref{FIG2} (a). At low external magnetic field and low temperature region, the magnetization is almost zero. When the magnetic field increases, the magnetization is enhanced and becomes spin-polarized as the saturation magnetization. The magnetic susceptibility as the fluctuation of magnetization is shown in FIG. \ref{FIG2} (b). The boundary of the magnetic phase diagram can be separated by considering the high magnetic susceptibility as the phase transition. The magnetic phase is checked by taking the snapshot of the Monte Carlo simulation at equilibrium. Consequently, we have found that at low temperature and low magnetic field, the magnetic phase is an incommensurate (IC), where the long range spin texture is broken. When the external magnetic field is intermediate, the skyrmion lattice (SkL) can be stabilized. When the external magnetic field is sufficient large, the spin polarized of ferromagnetic state can be formed. The snapshot of the skyrmion phase will be discussed in FIG. 3.
 
The magnetization versus magnetic field ($M$--$H$) curve of CrCl$_{3}$ and CrBr$_{3}$ are shown in FIG. \ref{FIG2} (c) and (d), respectively. The hysteresis loop of CrCl$_{3}$, which shows a two--triangle shape with symmetry to the origin, is quite different from the rectangle shape of Stoner--Wohlfarth model such as the CrBr$_{3}$ case. The increase in the magnetic field can enhance the magnetization and stabilize the FM phase. However, after getting the saturation magnetization, when the magnetic field decreases to close the zero, the magnetization decreases rapidly to zero. This phenomenon in the hysteresis loop is similar to the magnetization reversal process of the vortex. This peculiar behavior of the hysteresis loop will be a fingerprint of the realization in skyrmion phase in CrCl$_{3}$.

\begin{figure*}[htp] 
\centering
\includegraphics[width=17.6cm]{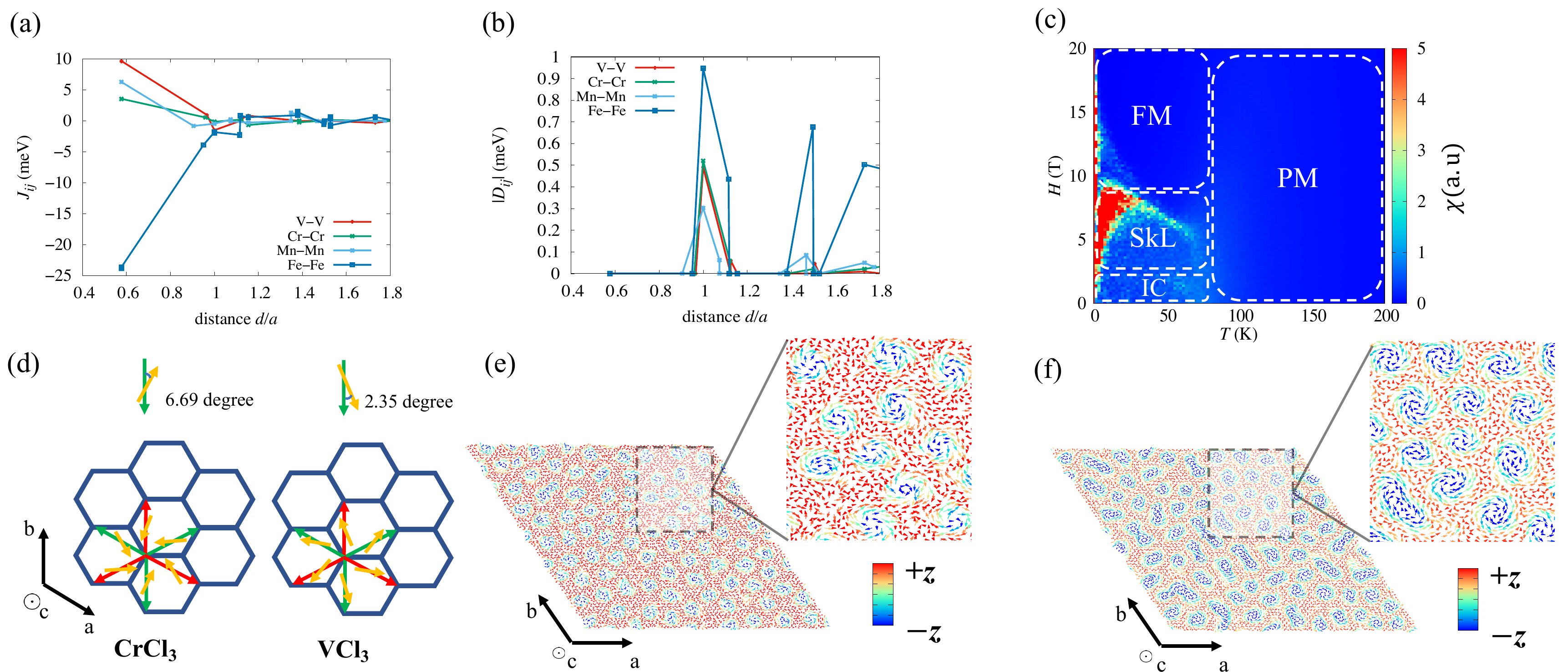} 
\caption{(a) The magnetic exchange coupling constants of M--M pairs of MCl$_{3}$ (M: V, Cr, Mn, Fe) as a function of distance over lattice constant ($d/a$). (b) The length of Dzyaloshinskii--Moriya vector of MCl$_{3}$ as a function of distance over lattice constant ($d/a$). (c) The magnetic susceptibility of VCl$_{3}$ as the function of temperature and the external magnetic field, where the magnetic phase is also specified. (d) The in--plane components of Dzyaloshinskii--Moriya vector for second nearest neighbor of intra--layer of CrCl$_{3}$ and VCl$_{3}$, where the red and green arrow for the pair with different sub--lattice, and the yellow arrow indicates the direction of Dzyaloshinskii--Moriya vector. (e) The snapshot of Monte Carlo simulation of CrCl$_{3}$ at 10 K and the external magnetic field equal to 4 T along the positive $z$ direction. (f) The snapshot of Monte Carlo simulation of VCl$_{3}$ at 10 K and the external magnetic field equal to 6 T along the positive $z$ direction. }
\label{FIG3}
\end{figure*}

We have found that the skyrmion phase also exists in other MCl$_{3}$ materials. The magnetic exchange coupling constants and the Dzyaloshinskii--Moriya length of MCl$_{3}$ as a function of distance over lattice constant are shown in FIG. \ref{FIG3} (a) and (b), respectively. The first nearest neighbor of magnetic exchange coupling constant of V--V, Cr--Cr, and Mn--Mn pairs are positive while it is negative for Fe--Fe pair. On the other hand, the Dzyaloshinskii--Moriya vectors of VCl$_{3}$, MnCl$_{3}$, and FeCl$_{3}$ are comparable with CrCl$_{3}$, which is relatively large. Replacing Cl by Br and I can change the magnetic exchange coupling constants and Dzyaloshinskii-Moriya lengths. For the detail of the magnetic exchange coupling constants and Dzyaloshinskii-Moriya lengths of the other compounds, see Figure 1 in Supplementary materials. The magnetocrystalline anisotropy energy of VCl$_{3}$, MnCl$_{3}$, and FeCl$_{3}$ are 0.023, $-$0.140, 0.075 (meV/f.u), respectively. Only MCl$_{3}$ compounds are suitable to form the skyrmion due to the small MAE compared with DMI. The high MAE in the other compounds suppresses the effect of DMI to form the skyrmion. See Table I in Supplementary materials for the detail of the MAE value in the other compounds. The magnetic susceptibility of VCl$_{3}$, which depends on the external magnetic field and temperature, are shown in FIG. \ref{FIG3} (c). VCl$_{3}$ with small magnetic anisotropy and relatively high DMI, which is important to stabilize the skyrmion lattice, gives similar magnetic phase diagram as CrCl$_{3}$ in FIG. \ref{FIG2} (b). It is noteworthy that the skyrmion lattice in VCl$_{3}$ can survive at much larger external magnetic field than CrCl$_{3}$ although the size of Dzyaloshinskii--Moriya length is similar.

The in--plane components of Dzyaloshinskii--Moriya vector for second nearest neighbor of intra--layer of CrCl$_{3}$ and VCl$_{3}$ are shown in FIG. \ref{FIG3} (d). It is clear that the in--plane direction of Dzyaloshinskii--Moriya vector of CrCl$_{3}$ and VCl$_{3}$ are nearly opposite. It leads to the different helicity of skyrmion with Bloch--type, which is an additional degree of freedom. Since the helicity of skyrmion is different when changing from CrCl$_{3}$ to VCl$_{3}$, it gives a possibility of controlling the helicity of skyrmion through electric--field driven electron and hole dope. The snapshot of the Monte Carlo simulation of CrCl$_{3}$ and VCl$_{3}$ are shown in FIG. \ref{FIG3} (e) and (f). The skyrmion of CrCl$_{3}$ is Bloch--type with clockwise due to the direction of Dzyaloshinskii--Moriya vector. On the other hand, the skyrmion of VCl$_{3}$ is also Bloch--type with the rotation of spin being anti--clockwise. Due to the non--vanishing of Dzyaloshinskii--Moriya vector at the second nearest neighbor of intra--layer, the diameter of the skyrmion of both CrCl$_{3}$ and VCl$_{3}$ are quite small (about 2 nm) as the properties of centrosymmetric material, which is similar to the value in other centrosymmetric material originating from frustration or RKKY interaction\cite{Kurumaji2019S,Khanh2020NN,Takagi2022NC}. The snapshots of the other compounds in Monte Carlo simulations are shown in Figure 3 in Supplementary materials.

\begin{figure}[htp] 
\centering
\includegraphics[width=8.6cm]{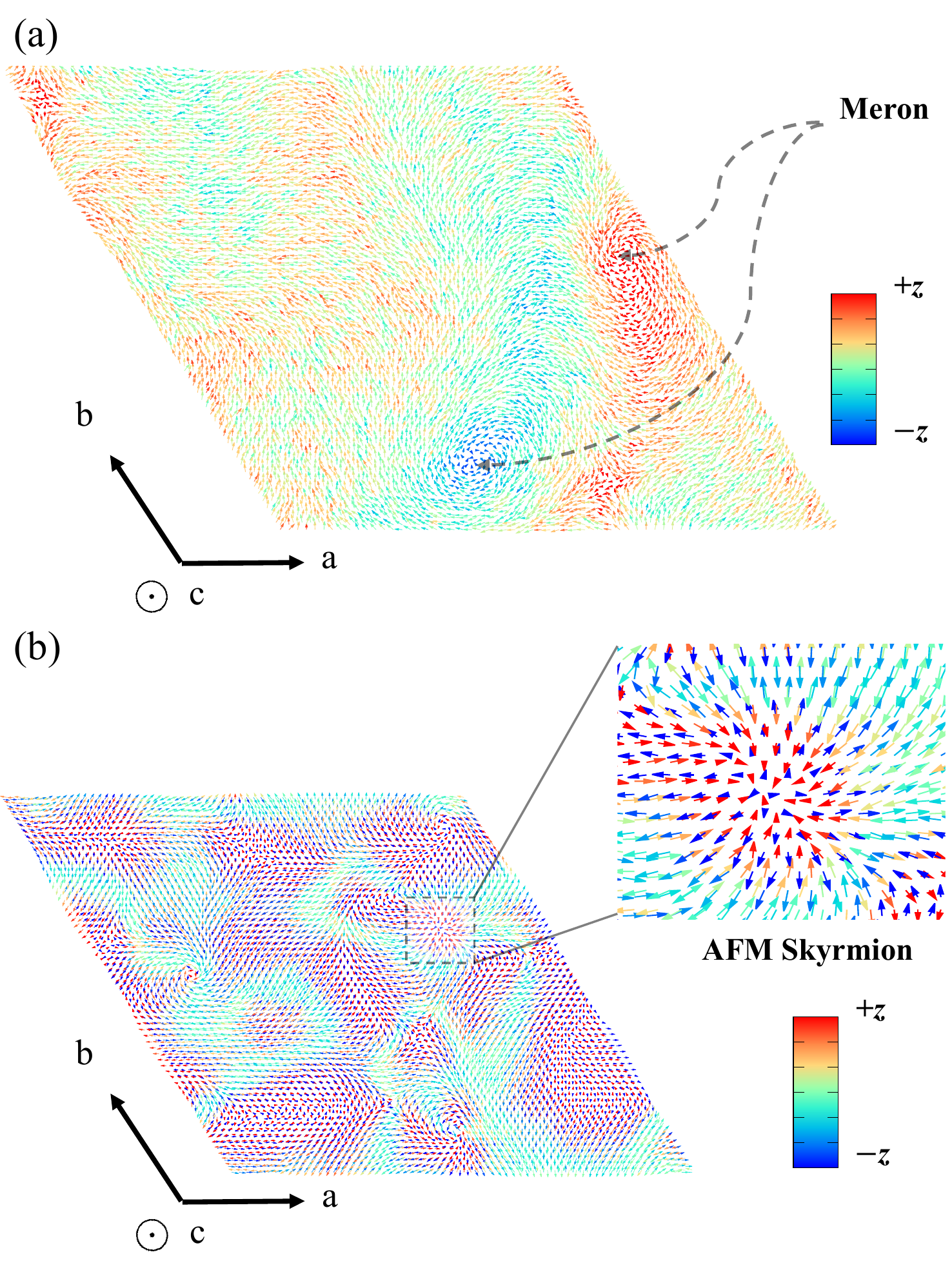} 
\caption{(a) The snapshot of Monte Carlo simulation of MnCl$_{3}$ at 10 K and the zero external magnetic field. (b) The snapshot of Monte Carlo simulation of FeCl$_{3}$ at 10 K and the external magnetic field equal to 4 T along the positive $z$ direction. } 
\label{FIG4}
\end{figure}

The snapshot of MnCl$_{3}$ at 10 K and zero external magnetic field is shown in FIG. \ref{FIG4} (a). The MAE value of MnCl$_{3}$ is $-$0.140 meV/(Mn--atom) as easy plane ($ab$ plane). In this case, meron, where the spin texture aligns in the in--plane compared with out--of--plane of skyrmion, can survive in MnCl$_{3}$ at zero external magnetic field due to the assistance of in--plane MAE. It means that meron in MnCl$_{3}$ does not require the external magnetic field as the skyrmion case, which might reduce the power consumption to maintain the skyrmion lattice as CrCl$_{3}$ and VCl$_{3}$.
 
The snapshot of FeCl$_{3}$ at 10 K and external magnetic field equal to 4 T along the positive $z$ direction are shown in FIG. \ref{FIG4} (b). FeCl$_{3}$ has negative sign in first and second nearest neighbor of intra--layer and interlayer, and also highest Dzyaloshinskii--Moriya interactions among them as the FIG. \ref{FIG3} (a) and (b). It is important to stabilize the antiferromagnetic (AFM) skyrmion, where the Hall effect is negligible compared with conventional skyrmion\cite{Zhang2016SR}. In this case, the AFM skyrmion is more convenient to be used in race track memory device without the movement to the boundary as Hall velocity in conventional skyrmion. The strong negative of first nearest neighbor can lead to the survivor of AFM skyrmion above room temperature. However, it also leads to a large size of AFM skyrmion compared with conventional skyrmion such as CrCl$_{3}$ and VCl$_{3}$.

In summary, the magnetic properties of van der Waals centrosymmetric MX$_{3}$ (M: V, Cr, Mn, Fe. X: Cl, Br, and I) materials are studied by combining first--principles calculations and Monte Carlo simulations. The effect of Dzyaloshinskii--Moriya interaction, which is usually ignored in centrosymmetric lattice, is considered in this study. The Dzyaloshinskii--Moriya interaction of second nearest neighbor intra--layer is non-negligible, which originates from the broken of local inversion symmetry. It becomes an indispensable factor to determine the magnetic properties of these compounds such as the skyrmion in CrCl$_{3}$ and VCl$_{3}$. The small size of skyrmion in the centrosymmetric materials is promising for spintronic applications such as high--density magnetic memory devices. Not only conventional skyrmion (CrCl$_{3}$ and VCl$_{3}$) but also antiferromagnetic skyrmion (FeCl$_{3}$) and meron in MnCl$_{3}$ are found in centrosymmetric materials of the present work. Our present work extends a new class of material as centrosymmetric lattice with rare earth free, where the skyrmion can be existed due to the finite Dzyaloshinskii--Moriya interaction.

\section*{Methods}
The relaxed crystal structure parameters and magnetocrystalline anisotropy energy were obtained using the VASP code with optB86b exchange functional as the van der Waals density functional (vdW-DF)\cite{Kresse1996PRB,Klimes2011PRB}. The magnetic exchange coupling constants and Dzyaloshinskii--Moriya vectors were calculated based on the Green function multiple-scattering formalism, as implemented in the SPR--KKR code\cite{Liechtenstein1987JMMM,EbertSPRKKR,Ebert2011RPP,Mankovsky2017PRB}. Monte Carlo simulations were performed using an in--house program with Metropolis algorithm for classical Heisenberg model\cite{HBT2022PRB,HBT2022AM}.

\section*{Data availability}
The datasets generated and/or analyzed during the current study are available from the corresponding author upon reasonable request.

\section*{Acknowledgements}
The authors would like to thank Prof. Tamio Oguchi and Dr. Kunihiko Yamauchi for the valuable discussions. This work was supported by MEXT via the ”Program for Promoting Researches on the Supercomputer Fugaku” (JP-MXP1020200205) and JSPS KAKENHI via the ”Grant-in-Aid for Scientific Research(A)” Grant Number 21H04553. The computation in this work was performed using the supercomputer Fugaku provided by the RIKEN Center for Computational Science, the Supercomputer Center at the Institute for Solid State Physics at the University of Tokyo, and the TSUBAME3.0 supercomputer at the Tokyo Institute of Technology.

\section*{Author Contributions}
H. B. T. developed the method and wrote the paper under the supervision of Y. M. All authors have approved the final version of the manuscript.

\section*{Competing interests}
The authors declare that they have no conflicts of interest.

\bibliography{basename of .bib file}

\begin{thebibliography}{99}
\bibitem{Fert2017NRM} Fert, A., Reyren, N., and Cros, V., Magnetic skyrmions: advances in physics and potential applications, {\it{Nat. Rev. Mater.}} 2, 17031 (2017).
\bibitem{Tokura2021CR} Tokura, Y., and Kanazawa, N., Magnetic Skyrmion Materials, {\it{Chem. Rev.}} 121, 2857-2897 (2021).
\bibitem{Moriya1960PR} Moriya, T., Anisotropic Superexchange Interaction and Weak Ferromagnetism, {\it{Phys. Rev.}} 120, 91 (1960).  
\bibitem{Kurumaji2019S} Kurumaji, T., et al., Skyrmion lattice with a giant topological Hall effect in a frustrated triangular-lattice magnet, {\it{Sci.}} 365, 914-918 (2019).
\bibitem{Khanh2020NN} Nguyen, D. K. et al., Nanometric square skyrmion lattice in a centrosymmetric tetragonal magnet, {\it{Nat. Nanotechnol.}} 15, 444-449 (2020).  
\bibitem{Takagi2022NC} Takagi, R. et. al., Square and rhombic lattices of magnetic skyrmions in a centrosymmetric binary compound, {\it{Nat. Comm.}} 13, 1472 (2022).  
\bibitem{Yambe2021SR} Yambe, R., and Hayami, S., Skyrmion crystals in centrosymmetric itinerant magnets without horizontal mirror plane, {\it{Sci. Rep.}} 11, 11184 (2021).
\bibitem{Yao2020NJP} Yao, X., Chen, J., and Dong, S., Controlling the helicity of magnetic skyrmions by electrical field in frustrated magnets, {\it{N. J. Phys.}} 22, 083032 (2020).
\bibitem{Psaroudaki2021PRL} Psaroudaki, C., and Panagopoulos, C., Skyrmion Qubits: A New Class of Quantum Logic Elements Based on Nanoscale Magnetization, {\it{Phys. Rev. Lett.}} 127, 067201 (2021).
\bibitem{Jing2022arxiv}Jing X. et al., Universal quantum computation based on nanoscale skyrmion helicity qubits in frustrated magnets, arXiv:2204.04589 (2022).
\bibitem{Song2020NE} Song, K. M., Skyrmion-based artificial synapses for neuromorphic computing, {\it{Nat. Electron.}} 3, 148–155 (2020). 
\bibitem{Ahn2020NPJ2D} Ahn, E. C., 2D materials for spintronic devices, {\it{NPJ 2D Mater. Appl.}} 4, 17 (2020).
\bibitem{Gibertini2019NN} Gibertini, M. et al., Magnetic 2D materials and heterostructures, {\i{Nat. Nanotechnol.}} 14, 408-419 (2019).
\bibitem{Cao2018N} Cao, Y. et al., Unconventional superconductivity in magic-angle graphene superlattices, {\it{Nature}} 556, 43–50 (2018).
\bibitem{Ando2006JPSJ} Ando, T., Screening Effect and Impurity Scattering in Monolayer Graphene. {\it{J. Phys. Soc. Jpn.}} 75, 074716 (2006). 
\bibitem{Koshino2018PRX} Koshino, M. et al., Maximally Localized Wannier Orbitals and the Extended Hubbard Model for Twisted Bilayer Graphene. {\it{Phys. Rev. X}} 8, 031087 (2018).
\bibitem{Nishi2017PRB} Nishi, H., Matsushita, Y., and Oshiyama, A., Band-unfolding approach to moir\'e-induced band-gap opening and Fermi level velocity reduction in twisted bilayer graphene. {\it{Phys. Rev. B}} 8, 085420 (2017).
\bibitem{Uchida2014PRB} Uchida, K. et al., Atomic corrugation and electron localization due to Moir\'e patterns in twisted bilayer graphenes, {\it{Phys. Rev. B}} 90, 155451 (2014).
\bibitem{Dresselhaus2010NL} Dresselhaus, M. S. et al., Perspectives on Carbon Nanotubes and Graphene Raman Spectroscopy, {\it{Nano Lett.}} 10, 751-758 (2010). 
\bibitem{McGuire2017C} McGuire, M. A., Crystal and Magnetic Structures in Layered, Transition Metal Dihalides and Trihalides, {\it{Cry.}} 7, 121 (2017).
\bibitem{Handy1952JACS} Handy, L. L., and Gregory, N. W., Structural properties of chromium(III) iodide and some chromium(III) mixed halides, {\it{J. Am. Chem. Soc.}} 74, 891-893 (1952).
\bibitem{HBT2022PRB} Tran., B. H. et al., Effect of magnetocrystalline anisotropy on magnetocaloric properties of an AlFe$_{2}$B$_{2}$ compound, {\it{Phys. Rev. B}} 105, 134402 (2022).
\bibitem{HBT2022AM} Tran., B. H. et al., Insight into anisotropic magnetocaloric effect of CrI$_{3}$, {\it{Acta Mater.}} 231, 117851 (2022).
\bibitem{Liechtenstein1987JMMM} Liechtenstein, A. I. at. al., Local spin density functional approach to the theory of exchange interactions in ferromagnetic metals and alloys, {\it{J. Magn. Magn. Mater.}} 67, 65--74 (1987).
\bibitem{EbertSPRKKR} Ebert, H. et al., The Munich SPR-KKR package, version 7.7, \url{https://www.ebert.cup.uni-muenchen.de/SPRKKR}.
\bibitem{Ebert2011RPP} Ebert, H., K\"odderitzsch, D., and Min\'ar, J., Calculating condensed matter properties using the KKR-Green's function method--recent developments and applications, {\it{Rep. Prog. Phys.}} 74, 096501 (2011).
\bibitem{McGuire2017PRM} McGuire, M. A. et al., Magnetic behavior and spin-lattice coupling in cleavable van der Waals layered CrCl$_{3}$ crystals, {\it{Phys. Rev. Mater.}} 1, 014001 (2017).
\bibitem{Kresse1996PRB} Kresse, G., and Furthm\"uller, J., Efficient iterative schemes for ab initio total--energy calculations using a plane--wave basis set, {\it{Phys. Rev. B}} 54, 11169--11186 (1996).
\bibitem{Mankovsky2017PRB} Mankovsky, S., and Ebert, H., Accurate scheme to calculate the interatomic Dzyaloshinskii--Moriya interaction parameters, {\it{Phys. Rev. B}} 96, 104416 (2017).
\bibitem{Chen2018PRX} Chen, L. et al., Topological Spin Excitations in Honeycomb Ferromagnet CrI$_{3}$, {\it{Phys. Rev. X}} 8, 041028 (2018).
\bibitem{Zhang2016SR} Zhang, X., Zhou, Y., and Ezawa, M., Antiferromagnetic Skyrmion: Stability, Creation and Manipulation. {\it{Sci. Rep.}} 6, 24795 (2016).
\bibitem{Klimes2011PRB} Klimeš, J., Bowler, D. R., Michaelides, A., Van der Waals density functionals applied to solids, {\it{Phys. Rev. B}} 83, 195131 (2011).




\end{thebibliography}

\section*{Supplementary Figures and Tables}

Magnetocrystalline anisotropy energy (MAE) of MX$_{3}$ is shown in Table. \ref{TAB1}. All compounds show that MAE increases when the atomic number of X element increases form Cl to I. However, the uniaxial anisotropy with easy axis as $c$-axis can be changed to easy plane $ab$-plane with negative MAE in the case of VX$_{3}$ and FeX$_{3}$. On the other hand, the high MAE of MX$_{3}$ (X=Br, I) suppresses the effect of Dzyaloshinskii-Moriya interactions compared with MCl$_{3}$ cases. The calculations of FeI$_{3}$ show that the material is non-magnetic (NM) with negligible local magnetic moment in Fe site.  

\begin{table}[!ht]
\centering
\caption{Magnetocrystalline anisotropy energy of MX$_{3}$ (M=V,Cr,Mn,Fe; X=Cl,Br,I)}    
\begin{tabular}{|l|l|l|l|l|}
\hline
        $E$$_{100}$-$E$$_{001}$ (meV/f.u) & M=V & M=Cr & M=Mn & M=Fe \\ \hline
       X=Cl & 0.023 & 0.034 & -0.140 & 0.075 \\ \hline
       X=Br & -0.156 & 0.187 & -1.321 & -0.539 \\ \hline
       X=I & -1.543 & 0.590 & -4.457 & ~ \\ \hline
\end{tabular}
\label{TAB1}
\end{table}

\begin{figure*}[htp] 
\centering
\includegraphics[width=16.6cm]{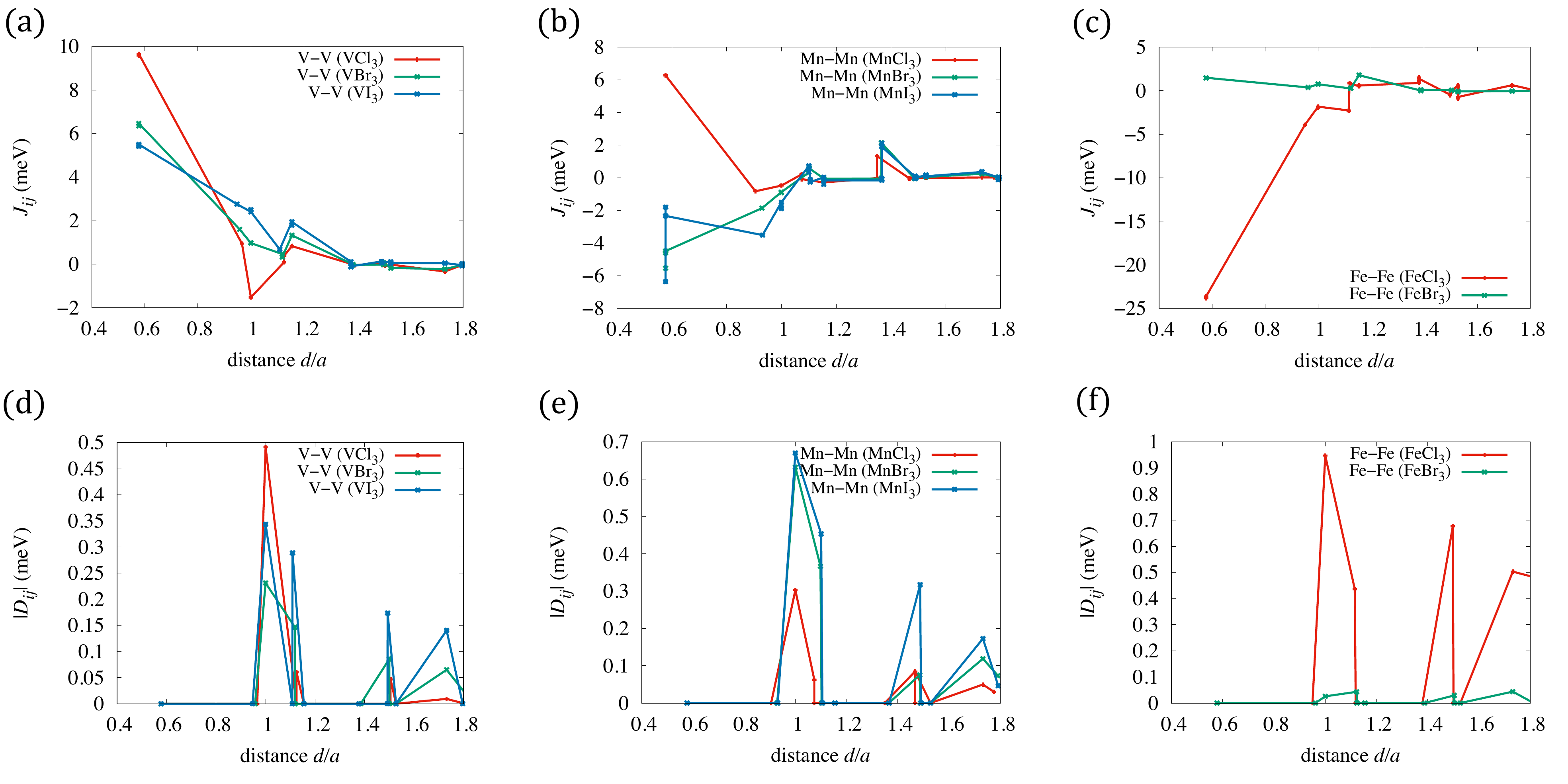} 
\caption{Isotropic magnetic exchange coupling constant as a function of distance over lattice constant $d/a$ of MX$_{3}$ (a), (b), and (c). Dzyaloshinskii-Moriya vector lengths as a function of distance over lattice constant $d/a$ of MX$_{3}$ (d), (e), and (f). } 
\label{FIG1_SUP}
\end{figure*}

The magnetic exchange coupling constants and Dzyaloshinskii-Moriya vector lengths as a function of distance over lattice constant $d/a$ of MX$_{3}$ (M=V, Mn, Fe;X=Cl, Br, I) are shown in FIG. \ref{FIG1_SUP}. The first nearest neighbor of magnetic exchange coupling constant in VBr$_{3}$ and VI$_{3}$ does not changed so much compared with VCl$_{3}$. It leads to ferromagnetic (FM) order in one layer with strong in-plane MAE. On the other hand, the negative magnetic exchange coupling constants in the case of MnBr$_{3}$ and MnI$_{3}$ lead to antiferromagnetic (AFM) order. The strong in-plane MAE in MnBr$_{3}$ and MnI$_{3}$ also suppresses the effect of Dzyaloshinskii-Moriya interactions. For the FeBr$_{3}$, the magnetic exchange coupling constants are positive but relatively small due to the small local magnetic moment at Fe site. It leads to the in-plane FM order in FeBr$_{3}$ case.

\begin{figure*}[htp] 
\centering
\includegraphics[width=16.6cm]{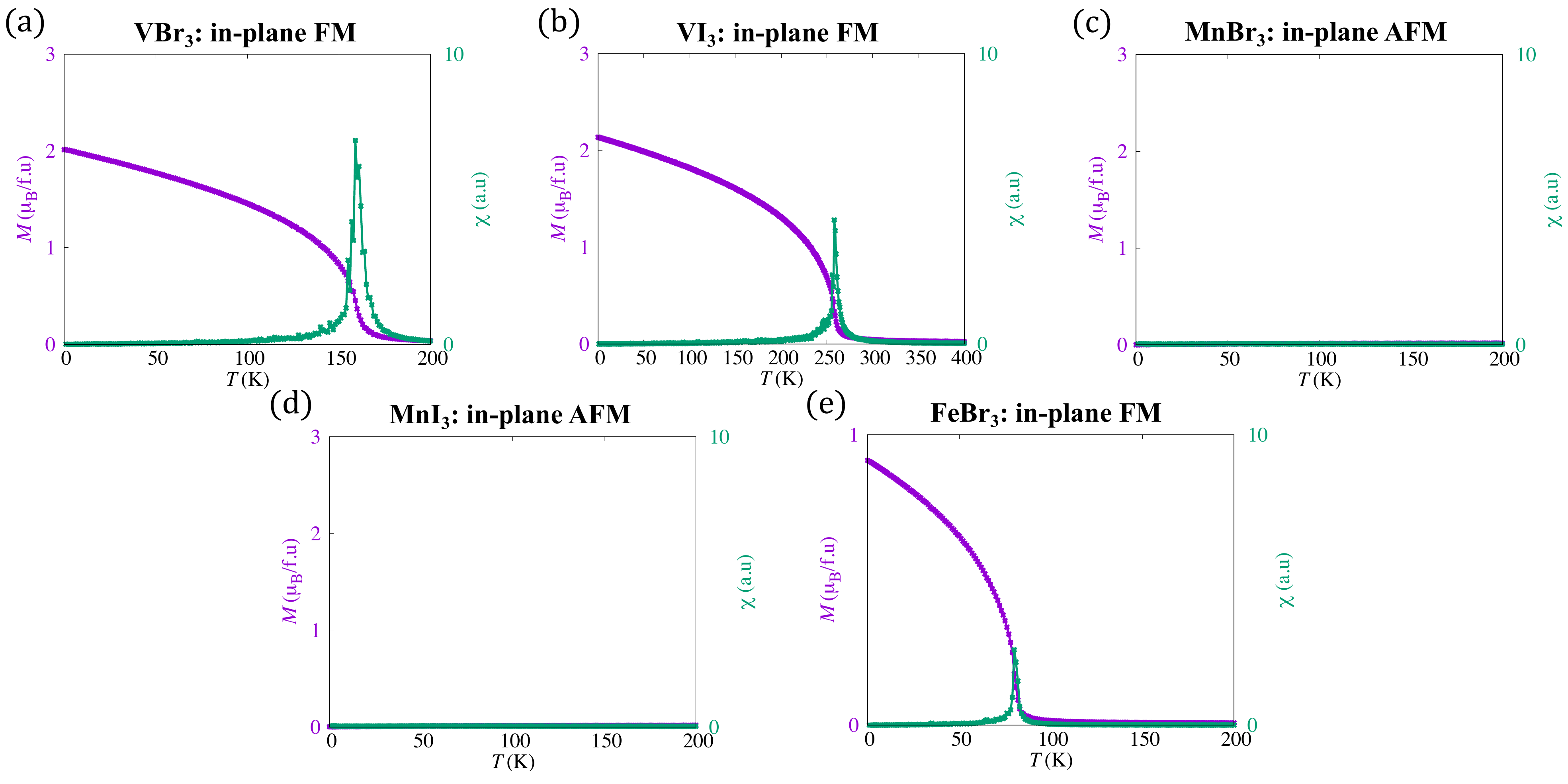} 
\caption{Magnetization (purple) and magnetic susceptibility (green) versus temperature of VBr$_{3}$ (a), VI$_{3}$ (b), MnBr$_{3}$ (c), MnI$_{3}$ (d), and FeBr$_{3}$ (f). } 
\label{FIG2_SUP}
\end{figure*}

The magnetization and magnetic susceptibility dependence on temperature of MX$_{3}$ (M=V, Mn, Fe;X=Br, I) are shown in FIG. \ref{FIG2_SUP}. The magnetization of VBr$_{3}$, VI$_{3}$, and FeBr$_{3}$ shows a clear ferromagnetic-paramagnetic (FM-PM) transition, where the Curie temperatures correspond to the divergence of magnetic susceptibility. On the other hand, the magnetization and magnetic susceptibility of MnBr$_{3}$ and MnI$_{3}$ are approximately equal to zero, where the ground states are AFM as the result of negative magnetic exchange coupling constants.

\begin{figure*}[htp] 
\centering
\includegraphics[width=16.6cm]{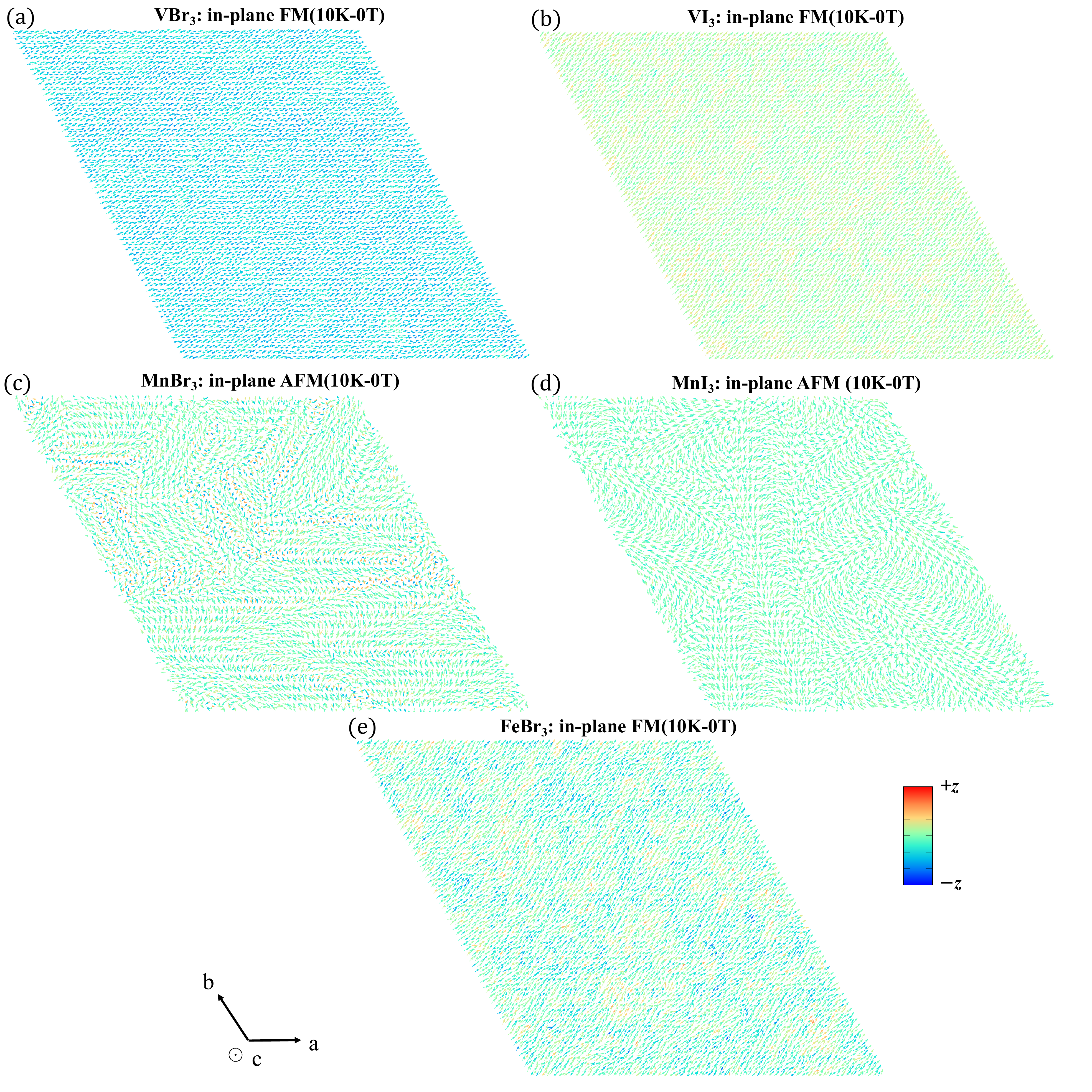} 
\caption{Snapshot of one layer at temperature being 10K and zero external magnetic field 0T of VBr$_{3}$ (a), VI$_{3}$ (b), MnBr$_{3}$ (c), MnI$_{3}$ (d), and FeBr$_{3}$ (f). } 
\label{FIG3_SUP}
\end{figure*}

The snapshot of one layer of simulation cell of MX$_{3}$ (M=V, Mn, Fe;X=Br, I) at temperature being 10K and zero external magnetic field 0T  are shown in FIG. \ref{FIG3_SUP}. From the snapshot, all compounds show in-plane magnetic favor due to the negative MAE. The configuration of VBr$_{3}$, VI$_{3}$, and FeBr$_{3}$ are ferromagnetic (FM) while the MnBr$_{3}$ and MnI$_{3}$ are antiferromagnetic (AFM).

\end{document}